\documentclass[conference]{IEEEtran}
\usepackage{amsmath,amsfonts}
\usepackage{array}
\usepackage[caption=false,font=normalsize,labelfont=sf,textfont=sf]{subfig}
\usepackage{textcomp}
\usepackage{stfloats}
\usepackage{url}
\usepackage{enumitem}
\usepackage{verbatim}
\usepackage{graphicx}
\usepackage{algorithm}
\usepackage{algpseudocode}

\usepackage{booktabs} 
\usepackage{stackengine}
\usepackage{float}
\usepackage{tikz}
\usepackage{xcolor}
\usepackage{wasysym}
\usepackage{colortbl}
\usepackage{soul}
\usepackage{multirow}
\usepackage{ragged2e}
\usepackage{hyperref}
\usepackage{comment}

\ifCLASSINFOpdf
\else
\fi
\begin{document}
\title{Opto-ViT: Architecting a Near-Sensor Region of Interest-Aware Vision Transformer Accelerator with Silicon Photonics \vspace{-0.5em}}

\author{Mehrdad Morsali$^{*,1}$, Chengwei Zhou$^{*,2}$, Deniz Najafi$^{*,1}$, Sreetama Sarkar$^{3}$, Pietro Mercati$^4$,\\ Navid Khoshavi$^{5}$, Peter Beerel$^{3}$, Mahdi Nikdast$^{6}$, Gourav Datta$^{2}$, and Shaahin Angizi$^1$ \vspace{0.3em}\\ \small $^1$New Jersey Institute of Technology, USA $^2$Case Western Reserve University, USA\\
$^3$University of Southern California, USA
$^4$Intel Corporation, USA $^5$AMD, USA $^6$Colorado State University, USA\\
$^{*}$ Equal Contributions\\
E-mails: shaahin.angizi@njit.edu, gourav.datta@case.edu, pabeerel@usc.edu, mahdi.nikdast@colostate.edu, \\\vspace{-1.5em}
 \vspace{-2em}
\\}

\maketitle
\begin{abstract}
Vision Transformers (ViTs) have emerged as a powerful architecture for computer vision tasks due to their ability to model long-range dependencies and global contextual relationships. However, their substantial compute and memory demands hinder efficient deployment in scenarios with strict energy and bandwidth limitations. In this work, we propose \textit{Opto-ViT}, the first near-sensor, region-aware ViT accelerator leveraging silicon photonics (SiPh) for real-time and energy-efficient vision processing. Opto-ViT features a hybrid electronic-photonic architecture, where the optical core handles compute-intensive matrix multiplications using Vertical-Cavity Surface-Emitting Lasers (VCSELs) and Microring Resonators (MRs), while nonlinear functions and normalization are executed electronically. To reduce redundant computation and patch processing, we introduce a lightweight Mask Generation Network (MGNet) that identifies regions of interest in the current frame and prunes irrelevant patches before ViT encoding. We further co-optimize the ViT backbone using quantization-aware training and matrix decomposition tailored for photonic constraints. Experiments across device fabrication, circuit and architecture co-design, to classification, detection, and video tasks demonstrate that Opto-ViT achieves 100.4 KFPS/W with up to 84\% energy savings with less than 1.6\% accuracy loss, while enabling scalable and efficient ViT deployment at the edge.
\end{abstract}

\IEEEpeerreviewmaketitle

\section{Introduction}
Despite the widespread adoption of the Internet of Things (IoT), it largely lacks embedded intelligence, relying predominantly on cloud-based decision-making. Recent efforts have focused on developing advanced CMOS image sensors capable of accelerating Deep Neural Networks (DNN) employed for computer vision (CV) workloads. One approach, Processing-Near-Sensor (PNS), integrates CMOS image sensors and processors on a single chip \cite{carey2013100,hsu20200,yamazaki20174,angizi2018cmp,morsali2023deep}. 
Another approach, Processing-In-Sensor (PIS), involves integrating computation units at the pixel level. In the PIS approach \cite{xu2020macsen,xu2021senputing,tabrizchi2023appcip,angizi2023pisa,abedin2022mr,tabrizchi2024pinsim,tabrizchi2024apris,najafi2024enabling}, data is processed before reaching the Analog-to-Digital Converter (pre-ADC), allowing transfer to on- or off-chip processors \cite{el1999pixel,song2022reconfigurable,roohi2023pipsim}. However, prior PNS/PIS research has primarily focused on simpler models like convolutional neural networks and multi-layer perceptrons, which, though successful for many CV applications, are now somewhat outdated. 

Meanwhile, transformer-based networks, after their remarkable success in sequence-to-sequence tasks in NLP \cite{attention}, are now starting to dominate the CV field \cite{Caron_2021_ICCV}. 
In particular, their self-attention mechanism enables the modeling of long-range dependencies and global context \cite{attention}, which are critical for complex vision tasks such as object detection, semantic segmentation, and fine-grained classification. Vision Transformers (ViTs), in particular, have demonstrated superior performance over traditional convolutional neural networks (CNNs) in many benchmarks. Their ability to scale model capacity and generalize across diverse datasets makes them an appealing choice for real-world edge deployments.
Although transformer-based models have achieved notable success, they incur significant computational resources, largely due to the high complexity of the self-attention score and projection calculations \cite{ahmed2025deepcompress}.
This limits their applicability and progress in resource-constrained systems, presenting key challenges such as long inference times and large memory requirements. 
To address these issues, various vision transformer accelerators have been proposed in recent years. An FPGA-based hardware accelerator architecture for vision transformer models is introduced in \cite{vita2023}. It is optimized for edge applications and presents a reasonable frame rate. In \cite{retransformer}, a memristor-based processing-in-memory transformer accelerator is proposed. It exploits the substantial parallelism provided by memristor crossbar arrays to accelerate the transformer inference operation. A silicon-photonics hardware accelerator for vision transformers has been proposed in \cite{tron2023afifi}. This accelerator uses a non-coherent architecture to implement the feedforward and MHA components of the Transformer architecture, employing MR banks to handle the necessary matrix multiplications. However, edge deployment of such accelerators has remained largely unexplored, primarily because \textit{(i)} Current architectures continue to struggle with energy-intensive ADC/DAC and peripherals, which, even when optimized \cite{choi2015energy,xu2020macsen,hsu2019ai}, remain power-hungry in sensing and computing tasks; \textit{(ii)} the integration of PIS/PNS introduces substantial area overhead and power consumption, often necessitating additional memory to store intermediate data \cite{angizi2023pisa,tabrizchi2023appcip,song2022reconfigurable,angizi2023near}; and \textit{(iii)} computational speed is limited by electronic systems (operating in the GHz range), which cannot inherently achieve the high speeds and extensive parallelism of optical systems, which approach photodetection rates exceeding 100 GHz \cite{sunny2021robin,najafi2025neuro,sunny2021crosslight,cheng2020silicon,morsali2024oisa}.

\noindent \textbf{Our Contributions.} In this work, we systematically explore, for the first time, the potential for edge deployment of vision transformers within the optical domain. We propose \textit{Opto-ViT}, the first near-sensor Vision Transformer (ViT) accelerator that leverages silicon photonics for real-time, energy-efficient vision processing. By integrating an optical core to offload computationally intensive and complex matrix multiplications and incorporating a lightweight region-of-interest (RoI) pruning network, Opto-ViT pushes the frontier of transformer acceleration to the edge. Our key contributions are:

\vspace{-0.15em}
\begin{itemize}[leftmargin=0.75em, itemsep=0pt, topsep=0pt]
    \item \textbf{First photonic accelerator for ViTs with region-aware processing:} We introduce a hybrid electronic-optic architecture that supports full ViT pipelines and selectively processes salient image regions using a compact RoI Mask Generation Network, enabling energy \& bandwidth savings at the edge.
    \item \textbf{Efficient optical MatMul with parallel VCSEL-MR design:} We design a photonic core that supports scalable matrix-matric multiplication (MatMul) operations via wavelength-division multiplexing and Vertical-Cavity Surface-Emitting Laser (VCSEL)-driven optical inputs, improving efficiency and reducing tuning delays.
    \item \textbf{Hardware-aware optimizations via matrix decomposition:} We propose a decomposed MatMul strategy that minimizes tuning bottlenecks, avoids intermediate buffering, and enables pipelined optical computation for high throughput.
    \item \textbf{Quantization-aware 
    \& ROI-aware ViT training pipeline:} We demonstrate robust accuracy under 8-bit quantization and introduce a region-of-interest (RoI) pruned ViT design that reduces patch-level computation with minimal accuracy loss.
    \item \textbf{Comprehensive cross-layer simulation framework:} We build a full-stack evaluation platform, from photonic devices to application-level metrics, that is validated using fabricated MRs and circuit-level models. Our platform yields up to 55\% energy savings with near-baseline accuracy across both image and video-based computer vision benchmarks.
\end{itemize}
\vspace{-0.5em}

\section{Background and Related Work}
\begin{figure}[t]
\begin{center}
\begin{tabular}{c}
\includegraphics [trim = 80 50 50 700, clip, width=\linewidth]{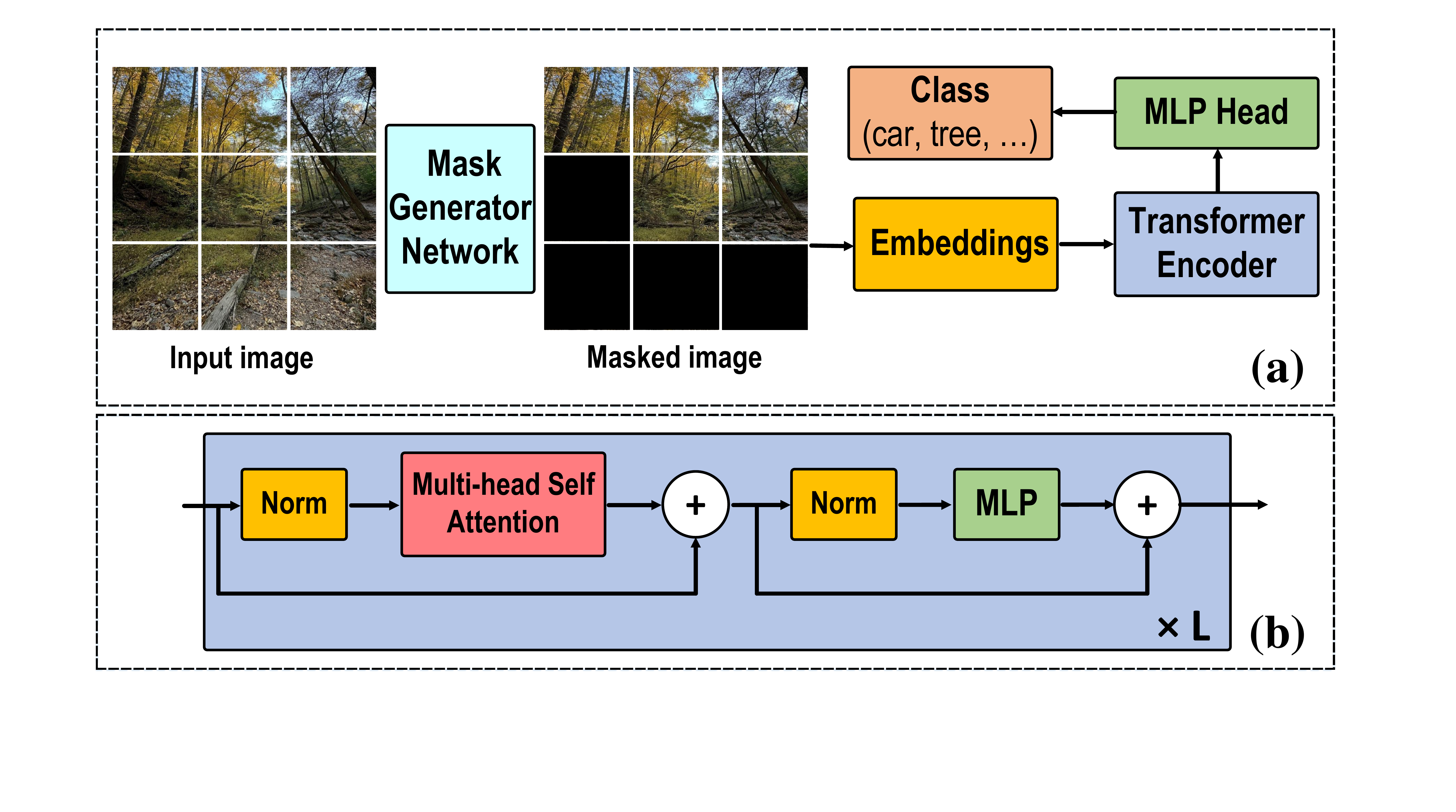}
\end{tabular} 
\vspace{-3em}
\caption{(a) Masked inference in Vision transformer (ViT), (b) Transformer encoder block in ViT.}
\label{transformer} \vspace{-1.5em}
\end{center}
\end{figure}

\noindent\textbf{Vision Transformers.}
The Vision Transformer (ViT) \cite{dosovitskiy2021an} is among the first to apply a transformer-based model architecture to vision tasks, closely resembling the encoder stack of the BERT transformer \cite{devlin2019bertpretrainingdeepbidirectional}. The encoder is structured as a cascade of $L$ encoder blocks. Fig. \ref{transformer} illustrates the architecture of the ViT model. In ViT, the input (\( X \)) is represented as a sequence of vectors, with each vector corresponding to a specific image patch. These vectors are produced by an embedding layer prior to the first encoder block. The input sequence [\( X_0 \), \( X_1 \), ...\( X_n \)] can be represented as a matrix with the size of \( n \)$\times$\( d_m \), where \( n \) is the number of patches, and \( d_m \) is the embedding dimension of the model. 
As shown in Fig. \ref{transformer}, each encoder block consists of two primary components: a Multi-Head Self-Attention (MHSA) module and a Feed-Forward Network (FFN) module. These components are typically organized with Layer Normalization (Norm) applied before each sub-layer, and residual connections that add the sub-layer's input to its output. The FFN module usually consists of two linear layers separated by a non-linear activation function, such as GELU. 
Each self-attention head 
includes three trainable weight matrices: the Query weight matrix ($W_Q$), the Key weight matrix ($W_K$), and the Value weight matrix ($W_V$), with the size of \( d_m \)$\times$\( d_k \), where \( d_k \) (the dimension of the weight vectors) can be calculated as $\frac{d_m}{h}$, and \( h \) is the number of heads. In each attention head, a set of query (\( Q \)), key (\( K \)), and value (\( V \)) matrices is derived by multiplying the input sequence \( X \) with the corresponding matrices \( W_Q \), \( W_K \), and \( W_V \). The self-attention output for each head is calculated by the scaled dot-product attention, defined as follows.
\vspace{-0.5em}
\begin{equation}
  \small  \text{Head}(X) = \text{attention}(Q, K, V) = \text{softmax}\left(\frac{Q K^T}{\sqrt{d_k}}\right)V
  \label{attention}
\end{equation}
 The output of the MHSA layer is generated by combining the outputs of all attention heads, which is then processed through the FFN module. This architecture facilitates the modeling of complex relationships within the input data while maintaining training stability and efficiency.

\begin{figure}[t] 
\centering
\includegraphics [width=0.99\linewidth,]{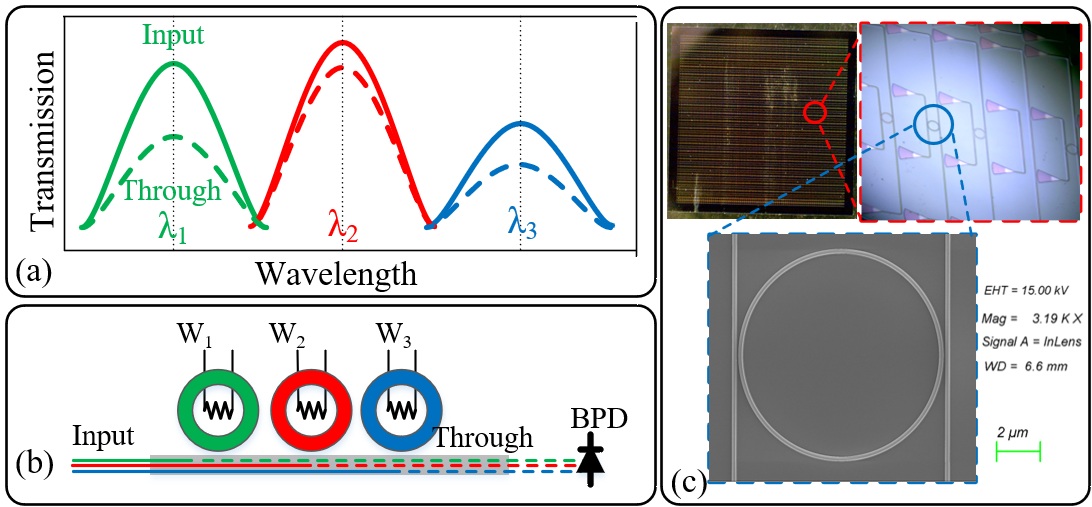}
\vspace{-1.8em}
\caption{(a) MR input and through ports’ spectra after imprinting a parameter (using tuning signal). By adjusting the MR's resonant wavelength , part of the input signal drops into the ring while the remaining propagates towards the through port, hence imprinting any parameter in the transmitted signals. (b) Multiple MRs in a single arm imprint weight values onto the input signal at different wavelengths. (c) Fabricated SiPh integrated circuit comprising of $>$200 identical copies of the MR cell layout (SEM image at the bottom).} 
\vspace{-1.5em}
\label{mr}
\end{figure}

\noindent\textbf{MRs and Photonics Acceleration.}
Silicon-photonic-based accelerators, known for their high bandwidth and innovative solutions to fan-in and fan-out challenges, offer considerable improvements in deep neural network (DNN) processing and machine vision applications \cite{sunny2021crosslight,liu2019holylight,zokaee2020lightbulb}. These accelerators are broadly classified into two types: Coherent designs use a single wavelength and encode parameters into the phase, the electric field amplitude, or the polarization of the optical signal \cite{zhao2019hardware}, and non-coherent designs, where they utilize multiple wavelengths to perform parallel operations and encode parameters within the amplitude of optical signals\cite{sunny2021crosslight,sunny2021robin}. In non-coherent systems, Microring Resonators (MRs) play a crucial role by dynamically selecting wavelengths and adjusting light intensity, serving as both inputs and weights \cite{sunny2021crosslight,sunny2021robin} or solely as weights \cite{morsali2024lightator}. Each MR has a resonant wavelength ($\lambda_{res}$) which can be computed by $\lambda_{res}=\frac{{n_{eff}}\times L}{m}$,
where $n_{eff}$ is the effective refractive index, $L$ is the MR's circumference, and $m$ is the resonant mode order \cite{bogaerts2012silicon}. The resonance peak of MRs can be shifted in wavelength by introducing small changes to the accumulated round-trip phase within the ring. This adjustment is typically accomplished through thermo-optic or electro-optic mechanisms. The MR can then modulate the amplitude of a passing light signal whose spectrum partially overlaps with the MR’s resonance spectrum, thereby applying a desired weight to light signals with wavelengths identical to the MR’s resonant wavelength. Fig. \ref{mr}(a) illustrates how the amplitude of input signals at specific wavelengths changes when passing through MRs tuned to the same resonance wavelength. As shown in Fig. \ref{mr}(b), if the MRs are tuned according to the weight values at their respective resonant wavelengths, and the activation values are applied as input light signals with matching wavelengths, the multiplication of weights and activations can be performed in parallel across different wavelengths known as Wavelength Division Multiplexing (WDM) method. At the end of the waveguide, a balanced photodetector (BPD) can accumulate the resulting optical signals and generate the Multiply-and-Accumulate (MAC) result. Prior work has demonstrated the potential of MR-based photonic approaches in DNN acceleration. LightBulb \cite{zokaee2020lightbulb}, for example, enhances binarized Convolutional Neural Networks (CNNs) with photonic XNOR operations. However, it suffers from high power consumption due to its dependence on Analog-to-Digital Converters (ADCs). ROBIN \cite{sunny2021robin} and CrossLight \cite{sunny2021crosslight} reduce bit-width for weights in convolutional layers, but their reliance on Digital-to-Analog Converters (DACs) and ADCs increases both the hardware footprint and power consumption. Recently, Lightator \cite{morsali2024lightator} was introduced as a near-sensor DNN accelerator capable of compressive frame acquisition and precise convolutional processing. However, to date, no silicon-photonic-based acceleration method has been developed specifically for vision transformers. Here we present our Opto-Vit accelerator, which utilizes the optical processing core for handling computationally heavy matrix-matrix multiplication (MatMul) operations of the vision transformer.

\section{Opto-ViT Architecture}

\subsection{Overview}
Vision Transformers (ViTs) are inherently computationally intensive due to the extensive MatMuls performed within each MHSA and FFN module across all encoder blocks. The primary operations involve MatMuls applied to sequences of input data. Variations among transformer models typically stem from differences in hyperparameters such as input size, patch size, embedding dimension, number of attention heads, and network depth. While FFN modules often account for a larger number of floating-point operations due to their expansive projection dimensions, MHSA modules present greater implementation challenges. This challenge arises because some matrices used in MHSA MatMul operations are intermediate results from previous computations. Managing these intermediate matrices—writing them to memory and reading them back—can introduce overhead that hinders computation and reduces processing speed. We discuss our strategy to minimize this overhead and enhance processing efficiency, in Section III-B below.

\begin{figure}[t] 
\centering
\includegraphics [width=1\linewidth,]{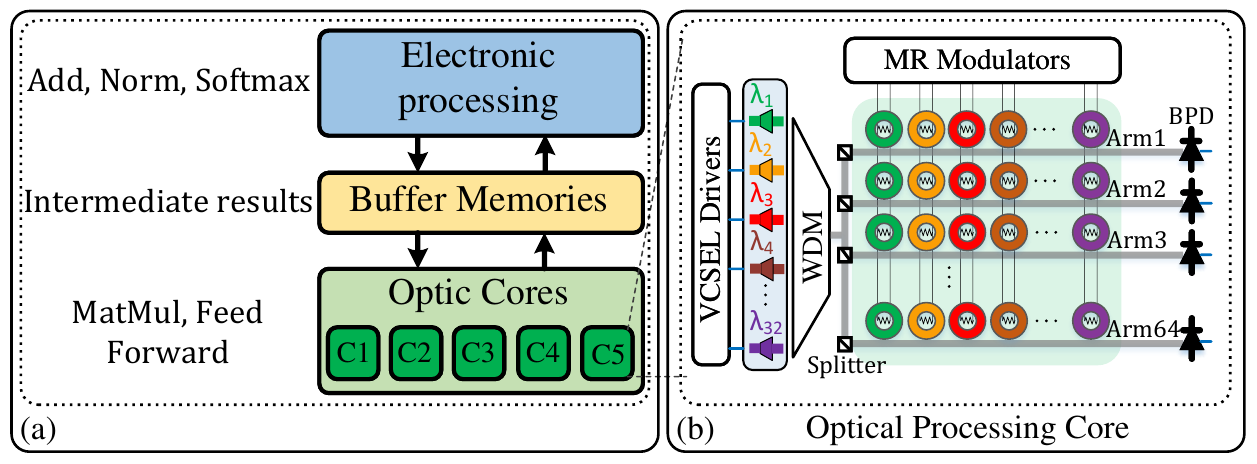}
\vspace{-2.1em}
\caption{(a) Overall architecture of Opto-ViT. (b) Architecture of an optic processing core in details } 
\vspace{-1em}
\label{arch}
\end{figure}

The proposed Opto-ViT accelerator architecture, depicted in Fig. \ref{arch}(a), adopts a hybrid electronic-optic scheme comprising two computational blocks and a buffer memory block. The primary computational component is the optical processing block, which houses five optical processing cores. This block handles the most computationally demanding tasks of the transformer, including MatMuls in the MHSA and FFN modules, and the embedding layer. Given that implementing non-linear functions such as Softmax and GELU is more efficient in the electronic domain, the electronic processing core is tasked with these operations, as well as normalization and the addition of intermediate results where necessary.

\noindent\textbf{Optical processing cores.} As discussed in the previous section, MAC operations can be implemented in the optical domain by tuning one set of operands on MRs and applying the other set as light signals. The amplitude of the light signals is modulated by the MRs, and the resulting signals are accumulated by BPDs. The processing principle of the optical core is based on this approach. 
As illustrated in the schematic shown in Fig. \ref{arch}(b), the core consists of MRs grouped into 32 wavelength channels along 64 waveguide arms (equal to \( d_k \) to facilitate matrix mapping and processing), which perform the multiplication; BPDs at the end of each arm, which handle the accumulation; an array of 32 VCSELs to generate input light signals with 32 distinct wavelengths; a multiplexer; VCSEL driver circuitry; and MR modulators for tuning the MRs. In our design, unlike most previous optical implementations, we directly use an array of VCSELs to generate light signals whose amplitudes represent the input data. This approach saves energy that would otherwise be spent tuning input values on MRs, as tuning MRs is generally more power- and time-consuming than driving VCSELs. Additionally, this method improves processing efficiency, as a single input light signal can be distributed to multiple arms and simultaneously multiplied by different weight values across those arms.
By leveraging the parallelism enabled by operating with multiple wavelengths and employing WDM, the optical core can efficiently perform Vector-Vector Multiplication (VVM). This capability can also be extended to MatMul by applying the rows of the input matrix as vectors to the core, while tuning the columns of the weight matrix onto the MRs in each arm of the core. In this way, each input row can be simultaneously multiplied by multiple columns. Depending on the matrix dimensions, the full MatMul operation can be completed over multiple cycles using repeated VVM. 

\noindent\textbf{Electronic processing unit.}
Thanks to the parallel processing capabilities enabled by multiple wavelengths and the extremely high computation speed of the optical domain, the optical core is well-suited for performing MAC operations. However, implementing the non-linear operations required by transformers, such as Softmax and GELU, is more challenging. Given the energy costs associated with optical-to-electrical and electrical-to-optical conversions, it is more efficient to perform these functions in the electrical domain. Therefore, an electronic processing unit has been incorporated, which includes a Softmax-GELU computation unit as proposed in \cite{softmaxgelu}, along with adder components to handle intermediate data addition during MatMUL operations where needed.

\noindent\textbf{Buffer memories.}
Buffer memories are designated for storing both the network’s weight values and the intermediate results generated by the optical core. They interface with the optical cores via DACs to load weight and intermediate values onto the tuning circuits, and via ADCs to retrieve the results produced by the BPDs. The size of the memory array is determined based on the specific application requirements. 

\subsection{Hardware Mapping \& Optimized processing flow}
\begin{figure}[b] \vspace{-1em}
\centering
\includegraphics [width=0.8\linewidth]{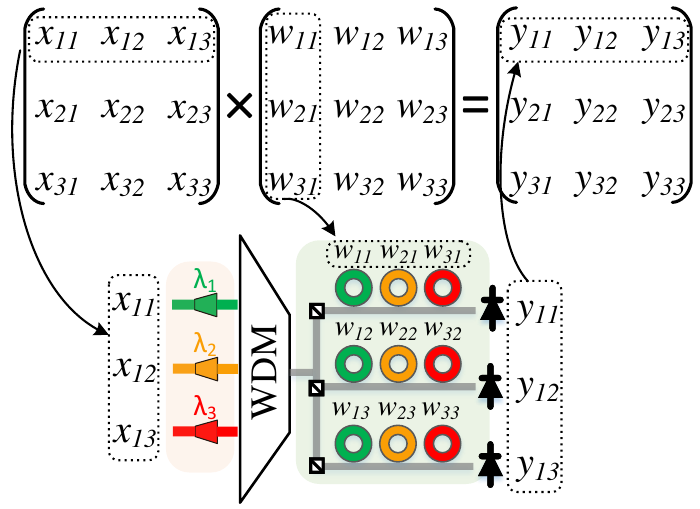}
\vspace{-1.2em}
\caption{Optical matrix-matrix multiplication } 
\vspace{-1.em}
\label{mapping}
\end{figure}

Fig. \ref{mapping} illustrates a simple example of performing a MatMul between two 3$\times$3 matrices in an optical core. As shown and previously described, the column elements of the second matrix, denoted as W, are tuned into the MRs within each arm of the core, with each arm corresponding to a column of the W matrix.
Next, the elements of the input matrix X are applied row by row as vectors to the VCSEL driver, modulating them into light intensities generated by a VCSEL array. These light signals, each carrying input data at different wavelengths, are then injected into the arms of the MR bank. As light passes through each arm, the MRs adjust the intensity at their resonance wavelengths, effectively computing dot products. At the end of each arm, a BPD accumulates these dot products, producing the final result of the MAC operation for that arm. Consequently, each BPD outputs one element of the resulting matrix per operational cycle. In the following cycle, the second row of matrix X is input to the VCSELs, and the same process computes the next row of the output matrix. This setup enables parallelism by feeding inputs into multiple arms and performing multiple MAC operations simultaneously, made possible through the use of different wavelengths, a capability facilitated by WDM in the optical domain.

The primary operation of the transformers is to compute the attention score described in equation \ref{attention}. To achieve this, we need to first calculate $Q.K^T$, where $Q$ and $K$ are obtained by multiplying the input $X$ with $W_Q$ and $W_K$, respectively. As mentioned, to perform MatMul of two matrices, we need to tune one of the participating matrices on MRs within the core. Thus, each MatMul requires a tuning step, which is time-consuming. Therefore, to calculate $Q.K^T$, we must first wait for the generation of $Q$ and $K$, then tune an MR bank by $K^T$, and finally perform the $Q.K^T$ multiplication. This sequence requires an additional tuning time for $K^T$, as we must wait until $K$ is generated. This approach not only disrupts the operational data flow and increases computation delay but also exacerbates the need for saving and buffering intermediate results from the MatMul operations. Consequently, it adds complexity and introduces additional tuning delays in the design. However, by decomposing matrices as proposed in \cite{retransformer}, we can eliminate the tuning wait time and alleviate issues related to intermediate data buffering and storage. This decomposition approach helps streamline the overall computation process. The matrix decomposition can be carried out as follows:
\begin{equation}
  \small   Q . K^T = Q . (X . W_K)^T = (Q . {W_K}^T ) . X^T . 
  \label{decomposition}
\end{equation}

\begin{figure}[t] 
\centering
\includegraphics [width=0.9\linewidth]{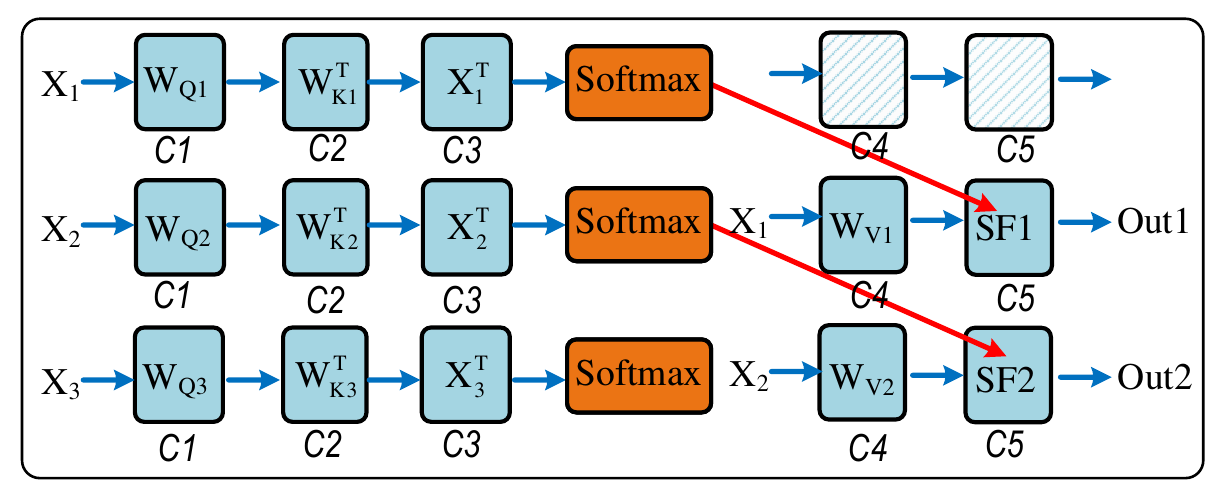}
\vspace{-1.2em}
\caption{Multi-core matrix-decompositional operational flow in attention heads  } 
\vspace{-1.em}
\label{pipe}
\end{figure}

\vspace{-0.7em}
Using this decomposition, we begin by tuning our MR banks with $W_Q$, ${W_K}^T$, and $X^T$, all available at the operation's start. This eliminates the need to wait for the intermediate result ($K$) to be generated. This operational flow eliminates one tuning step and removes the need to save and buffer intermediate values, enabling a pipelined operation and increasing the overall throughput of the design. Thus, as shown in Fig. \ref{pipe}, processing flow using five optical processing cores (C1 to C5) in our architecture starts with simultaneous tuning of cores C1 to C3 with $W_Q$, ${W_K}^T$, and $X^T$values, while C4 and C5 remains idle, waiting for the results to be generated and the Softmax computation to complete. In the next operational cycle, while C4 and C5 are being tuned by the softmax result and $W_V$ to compute the final output of the first input, C1 to C3 are tuned with the next set of operators. This pipelined approach ensures continuous input processing and effectively utilizes idle periods for tuning, which is one of the more time-consuming steps. It is worth mentioning that the result of MatMul operations must be scaled by the factor of $\frac{1}{\sqrt{d_k}}$ before the softmax operation. To prevent an extra division step, this scaling is applied to the weight parameters, and our weight MR bank is tuned by $\frac{{W_K}^T}{\sqrt{d_k}}$, directly. 

\begin{figure}[t]
\centering
\includegraphics [width=1.01\linewidth]{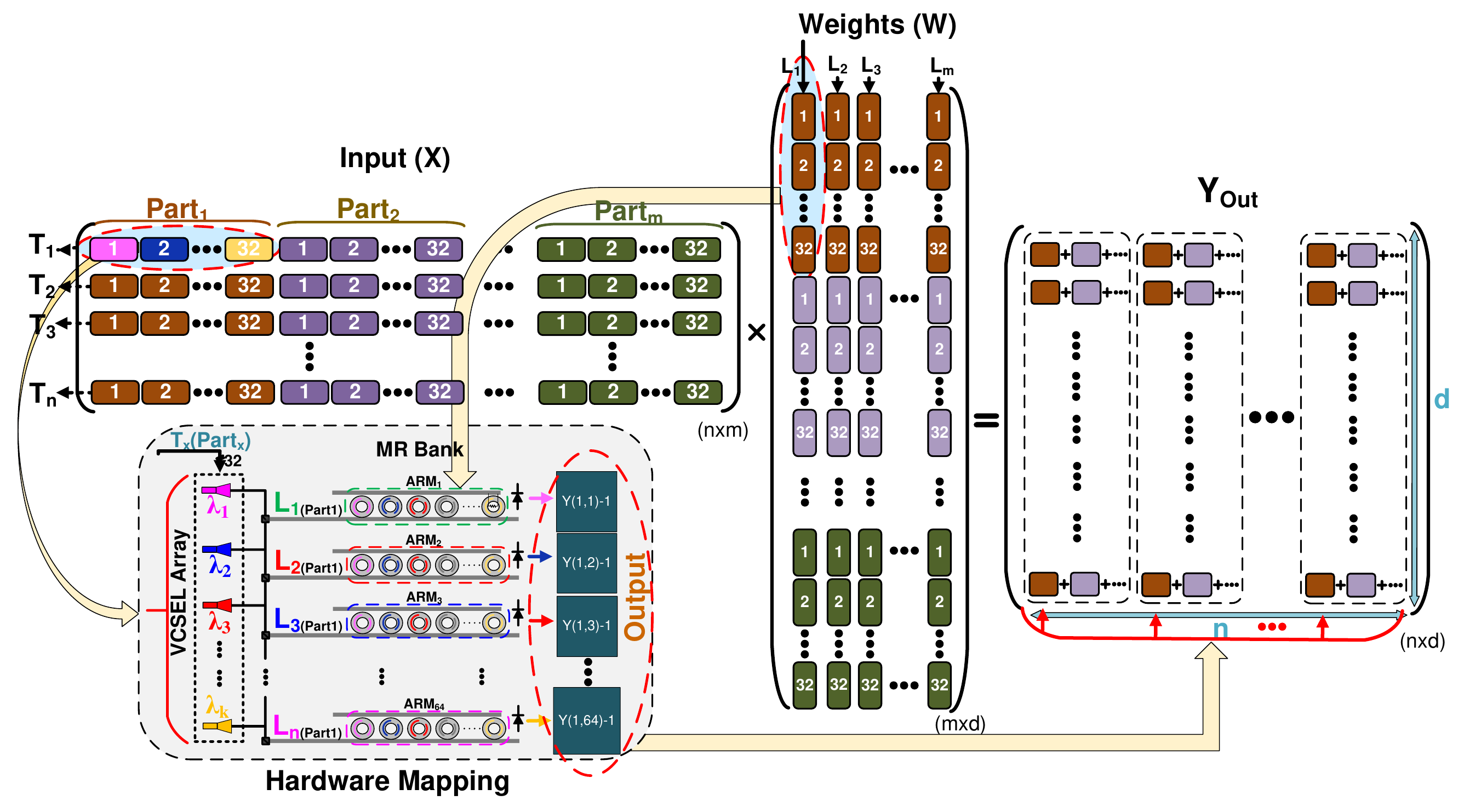}
\vspace{-2.1em}
\caption{Matrix splitting and hardware mapping methodology in Opto-ViT.}
\vspace{-0.6em}
\label{mapmethod}
\end{figure}
Another challenge in performing the MatMul operation lies in the large size of the weight matrices (\( d_m \)$\times$\( d_k \)), where \( d_m \) is typically ranges in the hundreds and \( d_k \) is often 64 in many transformer models. As a result, storing all weight elements within a single operational core is usually infeasible. To address this, the matrices must be partitioned into smaller chunks, and the multiplication is carried out incrementally over time. As shown in Fig. \ref{mapmethod}, input vectors are applied in segments to corresponding chunks of the weight matrices across different time slots. Based on our optical core architecture, we have 32 input wavelength channels available via VCSELs, enabling the generation of 32 input signals per cycle. These inputs are distributed across multiple arms and multiplied by corresponding weight values. In our design Each core contains 64 arms, matching the value of \( d_k \), Therefore, in each cycle, a chunk of 32 inputs can be applied to all columns of the weight matrix stored in the MRs, and the resulting intermediate values are stored. After all chunks of the input vector have been processed, the final matrix result is obtained by summing the corresponding intermediate results. Fig. \ref{mapmethod} illustrates this matrix mapping methodology using a color-coded scheme. This approach enables maximum utilization of matrix-level parallelism, fully leveraging the capabilities of the optical core.

\section{Experimental Results}

\begin{figure}[t] \vspace{-1em}
\centering
\includegraphics [width=0.99\linewidth]{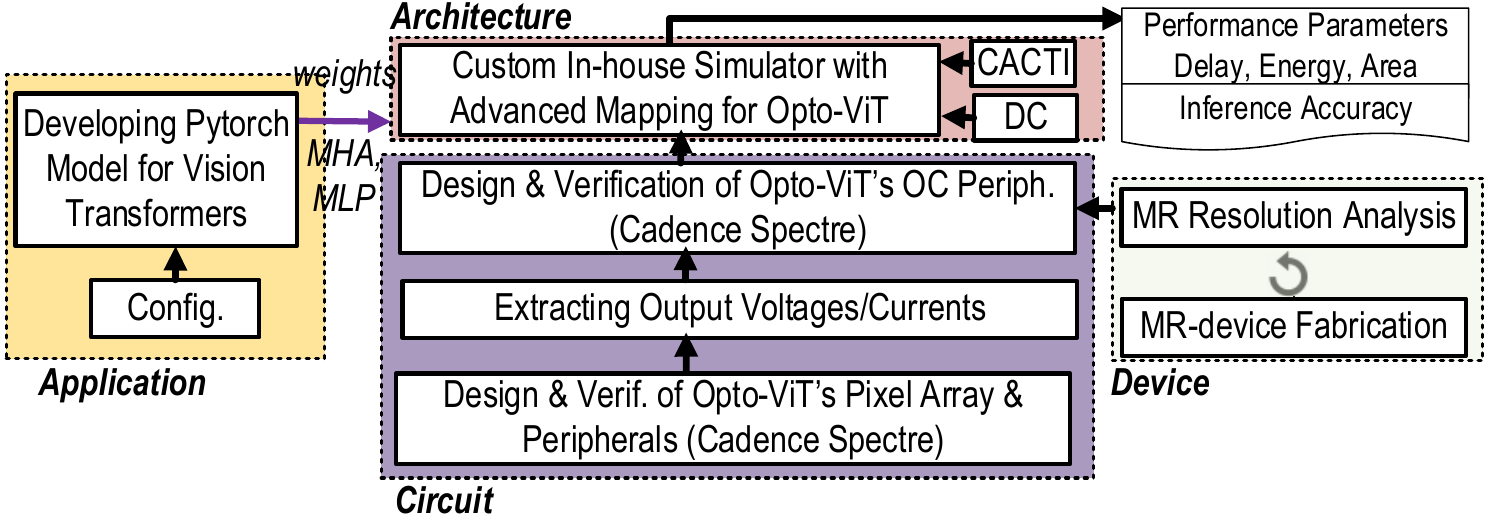}
\vspace{-2.1em}
\caption{Proposed bottom-up evaluation framework.}
\vspace{-1.1em}
\label{framework}
\end{figure}

\noindent\textbf{Setup.}
We developed a comprehensive, bottom-up experimental framework encompassing device, circuit, architecture, and application levels, as shown in Fig. \ref{framework}. At the device level, MR devices were fabricated and precisely calibrated to achieve 8-bit precision. 
To account for the impact of inevitable variations (e.g., fabrication-process variations) on the performance of the MR devices, $>$200 identical copies of the designed MR were placed on a 10$\times$10 mm$^2$ chip (see Fig. 2(c)), all of which were automatically measured and analyzed.
The resulting measurement data were then modeled and co-simulated with interface CMOS circuits using Cadence Spectre. Progressing to the circuit level, the pixel array and peripheral circuits were initially implemented in the 45nm NCSU Product Development Kit (PDK) library \cite{NCSU_PDK} within Cadence, where we obtained output voltages and currents. Subsequently, all components of the Opto-ViT, excluding memories, were developed in Cadence Spectre and Synopsys DesignCompiler \cite{DC}. At the application level, we developed and optimized four different variants of the ViT model in PyTorch. We then extracted the corresponding weight parameters for the MHA and MLP layers, which were then quantized and mapped onto the optical core to control the MR elements.

\noindent\textbf{MR Resolution Analysis.}
To achieve 8-bit MR resolution, MR designs have been explored by analyzing the mutual interference that occurs between optical signals in adjacent MRs due to crosstalk. Based on \cite{duong2014case}, we quantified the noise influence of one MR on another using the formula 
$\phi(i, j) = \frac{\delta^2}{(\lambda_i - \lambda_j)^2 + \delta^2}$, where, \(\phi(i - j)\) represents the noise from the \(j^{th}\) MR in the signal of the \(i^{th}\) MR, \((\lambda_i - \lambda_j)\) is the resonant wavelength difference, \((\lambda_i, \lambda_j)\), and \(\delta = \frac{\lambda}{2 \cdot Q_{factor}}\).  
A higher Q-factor, which corresponds to a sharper resonance, increases the susceptibility of MRs to noise; even minor shifts in the central frequency can lead to substantial signal loss, thereby constraining the achievable resolution. As a result, lower Q-factors are typically preferred. However, reducing the Q-factor involves trade-offs, such as increased device size and higher optical crosstalk, which can cause additional losses and increase tuning power requirements. The noise power component is calculated as
$P_{\text{noise}} = \sum_{i=1}^{n-1} \phi(i,j) P_{\text{in}}[i]$. For an input power intensity $P_{\text{in}}$ of 1, the resolution is determined by $\text{Resolution} = \frac{1}{\max|P_{\text{noise}}|}$.
Our analysis indicates that achieving at least 8-bit resolution requires MRs with a Q-factor of about 5000, along with tolerance to Fabrication Process Variations (FPVs). The Q-factor is sensitive to losses and dimensional variations in the MR. By selecting an input waveguide width of 400 nm, a ring waveguide width of 760 nm, and a radius of 5 $\mu$m, we improve tolerance to FPVs, maintain the desired Q-factor, and reduce the area footprint. This MR design, with a Q-factor of 5000, enables clear bit differentiation through slight intensity modulation, supporting accurate signal detection at the output port.

\noindent\textbf{Region of Interest Selection.}
We perform region of interest (RoI) selection solely based on the current input frame using a lightweight region \underline{M}ask \underline{G}enerator \underline{Net}work (MGNet) \cite{kaiser2024energy}. MGNet identifies semantically relevant input regions per frame and generates patch-wise binary masks. These masks are applied directly to the input (prior to the first ViT encoder block), enabling Opto-ViT to skip irrelevant patches early and reduce both optical compute and memory interface load. MGNet is composed of a single transformer block followed by a self-attention layer and a linear projection layer. It processes input images by dividing them into non-overlapping $\text{p}{\times}\text{p}$ patches and embedding each patch into a vector of length $L$. These embeddings are passed through the transformer block and self-attention module to compute attention scores for each patch. Specifically, MGNet computes an attention score $\text{S}_{{cls\_attn}}$ as the dot product between the query vector derived from the $cls\_token$ (${q}_{{class}}$) and the key matrix $K$ formed from the remaining patch embeddings.

\vspace{-3mm}
\begin{equation}
    {\text{S}_{cls\_attn}} = \frac{{q}_{class} {K}^T}{\sqrt{d}}.
    \label{eq:stoken}
\end{equation}

The attention score $\text{S}_{cls\_attn}$ inherently reflects the relative importance of each patch. This score is passed through a linear layer with an output dimension equal to the number of image patches, yielding region-wise or patchwise importance scores denoted as $\text{S}_{region}$. These scores are then passed through a sigmoid activation and thresholded using a region threshold $t_{reg}$ to produce a binary 2D mask, which we refer to as the input mask. MGNet is trained using a binary cross-entropy loss between the predicted region scores $\text{S}_{region}$ and the ground truth labels derived from the bounding boxes in each frame. In the ground truth label, a region is assigned a value of \textit{one} if it contains an object either fully or partially, and \textit{zero} otherwise. The accuracy of the generated mask is evaluated using  Intersection over Union (mIoU) between the predicted mask and the ground truth.

Unlike CNNs, where features are hierarchically extracted and spatially mixed across layers, ViTs operate on independent input patches throughout the network. This makes RoI-based patch pruning particularly effective: once a patch is masked at the input, all subsequent computations related to that patch are entirely skipped, yielding linear energy and compute savings. This property makes ViTs especially amenable to our region-aware acceleration approach compared to traditional CNNs.

\noindent\textbf{Performance Estimation.}
\begin{figure}[t] \vspace{-1em}
\centering
\includegraphics [width=0.97\linewidth]{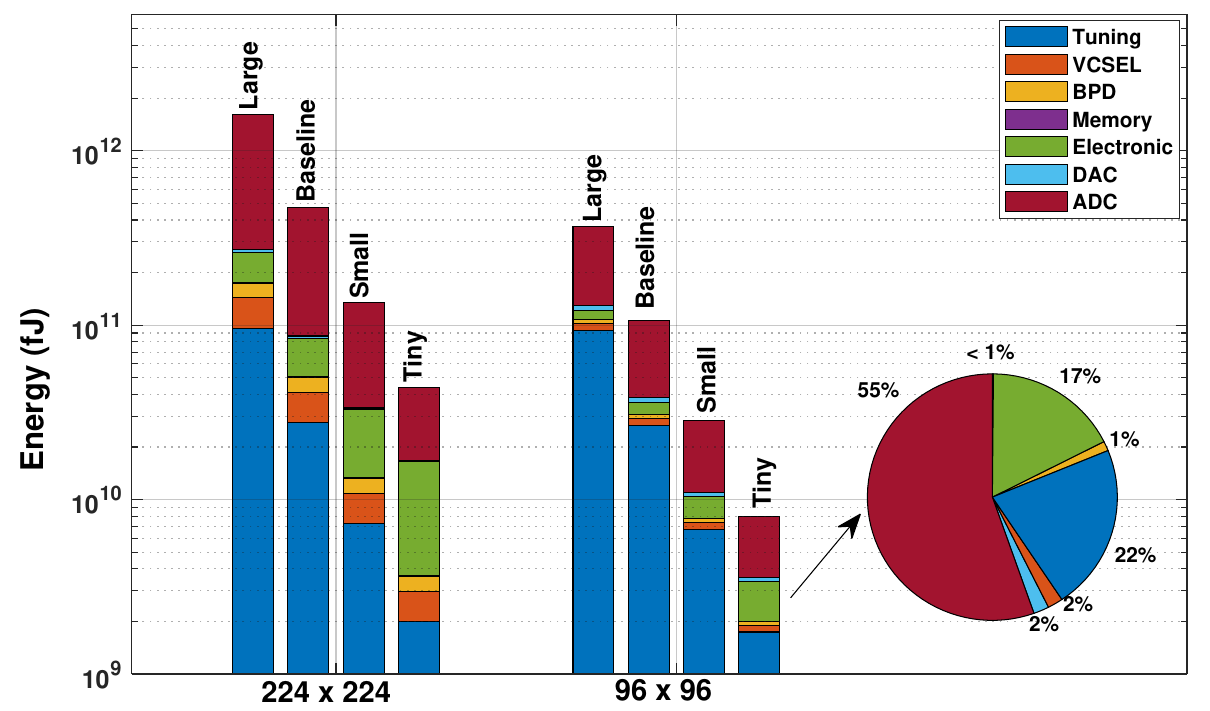}
\vspace{-1.0em}
\caption{Breakdown of energy consumption for processing different ViT models with two input image sizes: 224$\times$224 and 96$\times$96.}
\vspace{0.1em}
\label{energy1}
\end{figure}
To evaluate the performance parameters of the proposed design, we analyzed the energy consumption and processing delay of the accelerator while processing four different transformer networks (Large, Baseline, Small, and Tiny) using two different input image sizes: 224$\times$224 and 96$\times$96. Fig. \ref{energy1} shows the breakdown of energy consumption components, including Tuning, VCSEL, BPD, ADC, DAC, memory, and the electronic processing unit. A clear trend of energy reduction is observed when smaller networks and smaller input images are processed. Since the figure uses a logarithmic scale for better readability, the pie chart highlights the percentage contribution of each energy component in the design for the Tiny-96$\times$96 case. Although the main processing is performed in the optical domain using analog techniques, according to the pie chart, the ADCs still account for the largest share of energy consumption. This emphasizes the importance of further shifting processing toward the analog domain to reduce the overhead caused by data conversion.
The processing latency results are presented in Fig. \ref{delay1}, which shows the breakdown of delay components, including optical processing delay (with ADC and DAC delays included), electronic processing unit delay, and memory latency. As illustrated in the pie chart for the Tiny-96$\times$96 case, the optical processing stage contributes the most to the overall latency, as it handles the majority of the transformer’s computational workload. It can also be observed that memory latency exceeds the processing delay of the electronic unit, underscoring the importance of developing techniques to reduce intermediate data during transformer computations.
\begin{figure}[t] \vspace{-1em}
\centering
\includegraphics [width=0.99\linewidth]{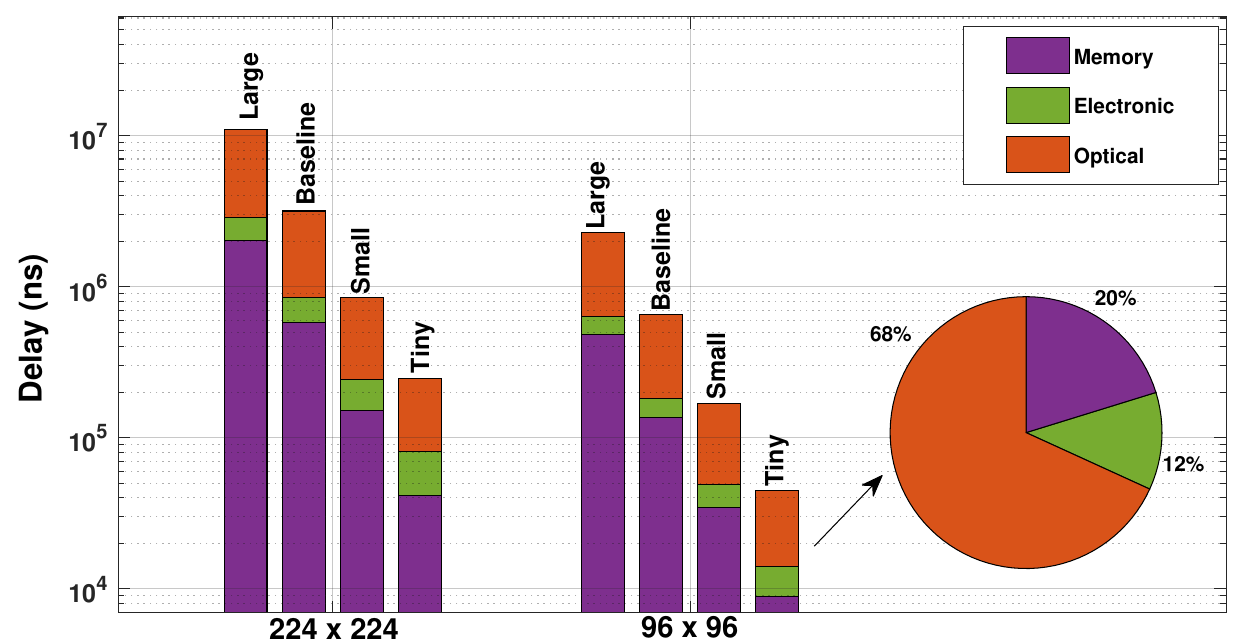}
\vspace{-2.1em}
\caption{Processing delay breakdown for various ViT models processed with two different input image sizes: 224$\times$224 and 96$\times$96.}
\vspace{-0.8em}
\label{delay1}
\end{figure}
In addition to the performance analysis of standard ViT models with varying configurations and image sizes, energy and latency evaluations have been conducted to demonstrate the efficiency of the ROI-based method. As explained above, a lightweight ViT network, referred to as MGNet, is employed before the backbone ViT for ROI selection. By skipping irrelevant patches, MGNet allows the backbone ViT to process a reduced number of patches, leading to significant energy savings and latency reduction. Fig. \ref{energyROI} illustrates the energy consumption of the baseline ViT used as a backbone network, processing different input image sizes. Two scenarios are shown: one where the backbone processes the full image with all patches, and another where it works in conjunction with the MGNet, which filters out irrelevant patches. Despite introducing a small additional energy overhead, using the MGNet significantly reduces overall energy consumption by decreasing the number of patches processed by the backbone. 
Examples of the number of RoI patches selected by the ViT masking algorithm for both 224$\times$224 and 96$\times$96 input image sizes are shown in Fig. \ref{energyROI}, along with their corresponding energy savings. 
The amount of energy saved is directly related to the number of ROI patches selected by the MGNet, thus, fewer ROI patches lead to greater energy efficiency.  
\begin{figure}[t] \vspace{-1em}
\centering
\includegraphics [width=0.99\linewidth]{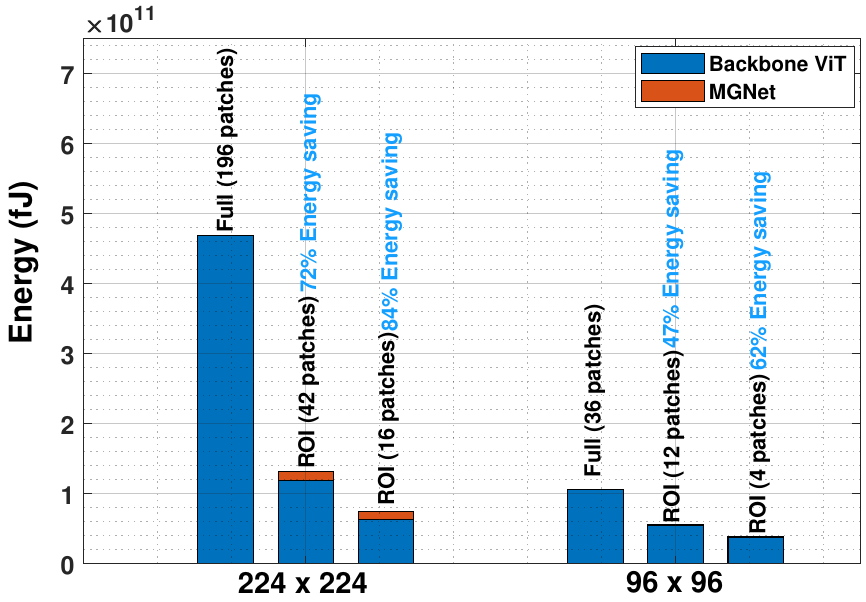}
\vspace{-2.1em}
\caption{Energy consumption of the accelerator processing baseline ViT with and without MGNet for ROI selection for two different input image sizes: 224$\times$224 and 96$\times$96.}
\vspace{-0.2em}
\label{energyROI}
\end{figure}
Fig. \ref{delayROI} also demonstrates the effectiveness of using the MGNet for ROI selection in reducing transformer processing latency, evaluated under the same conditions as those used for the energy analysis. A similar trend of latency reduction can be observed, with slightly greater improvements when the MGNet is applied prior to the backbone ViT. \\
\noindent\textbf{Accuracy Analysis.}
To enable deployment on optical neural network hardware—where device-level constraints such as photonic precision and ADC/DAC resolution limit numerical representation—we adopt 8-bit quantization for all Opto-ViT models. This strikes a balance between efficiency and accuracy while aligning with the precision limits of optical systems. Additionally, we constrain activation and weight dimensions to comply with the strict compute and area budgets of the photonic architecture. To preserve model performance under 8-bit quantization, we apply quantization-aware training (QAT) \cite{jacob2018quantization}. Unlike post-training quantization, QAT introduces quantization effects during training, allowing the model to gradually adapt to quantization artifacts. This is crucial for maintaining downstream task performance, especially in complex vision tasks like detection and segmentation.
Specifically, we leverage the straight-through estimator (STE) \cite{bengio2013estimating} to bypass the non-differentiability of quantization operations during backpropagation. Symmetric uniform quantization \cite{li2023vit} is used, with dynamic adjustment of the quantization range based on the statistics of model outputs. During training, quantized outputs are de-quantized to enable gradient-based optimization while faithfully simulating low-precision inference behavior. QAT is applied to both weights and activations in the patch‑embedding, MHSA, and FFN modules. 
In addition to quantization, we apply the lightweight MGNet described above, which enables early patch pruning by producing binary masks based on semantic relevance. For our implementation, MGNet uses patch size of 16, embedding dimension of 192, and 3 attention heads. 

\begin{figure}[t] \vspace{-1em}
\centering
\includegraphics [width=0.99\linewidth]{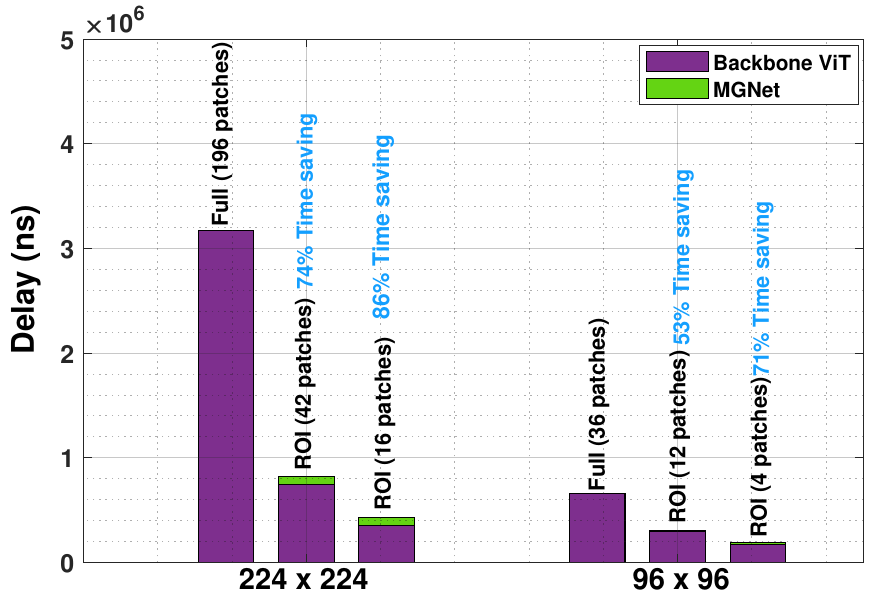}
\vspace{-2.1em}
\caption{Processing latency of the accelerator when running the baseline ViT model with and without MGNet for ROI selection, evaluated for two different input image sizes: 224$\times$224 and 96$\times$96.}
\vspace{-1.1em}
\label{delayROI}
\end{figure}

\noindent\textit{1) Image Classification:}
We fine-tuned various scales of ViT models (pre-trained on ImageNet-21k~\cite{ridnik2021imagenet} with patch size$=$16) for 100 epochs using SGD (learning rate$=$1e$-$2) on CIFAR-10~\cite{CIFAR10} and Tiny-ImageNet~\cite{deng2009imagenet}. CIFAR-10 images are resized to \(96\times96\), while Tiny-ImageNet inputs are evaluated at both \(96\times96\) and \(224\times224\), to study the impact of input resolution and RoI masking. Our results are summarized in Table~\ref{table:classification}. Compared to full-precision ViT baselines, the Opto-ViT models, quantized to 8-bit precision, achieve highly competitive performance. For instance, Opto-ViT-S reaches 97.67\% accuracy on CIFAR-10 at \(96\times96\), which is on par with the full-precision baseline ViT-B (97.86\%). Very subtle accuracy drop are observed across Tiny (Opto-ViT-T), Base (Opto-ViT-B), and Large (Opto-ViT-L) variants, confirming the robustness of QAT across model scales. On Tiny-ImageNet, increasing the input resolution from \(96\times96\) to \(224\times224\) yields a substantial improvement for Opto-ViT-B (from 80.56\% to 84.64\%). However, the introduction of an RoI mask (Opto-ViT-B Mask) at \(224\times224\) results in a noticeable drop in accuracy (down to 80.12\%, a 4.5\% decrease) while bring skip\% (pixel) of 67\% for input images. This degradation is primarily attributed to the lack of ground-truth masks or bounding box annotations in Tiny-ImageNet, which prevents proper training of the MGNet. Instead, the mask used here is transferred directly from the MGNet pre-trained on ImageNet VID, which introduces a domain mismatch and results in suboptimal patch pruning.

\noindent\textit{2) Object Detection and Segmentation:}
We conducted object detection and instance segmentation experiments on the COCO dataset \cite{lin2014microsoft} using a ViTDet base model \cite{li2022exploring} integrated into the Mask R-CNN framework \cite{he2017mask}. All models are trained for 20 epochs on input images resized and padded to 224$\times$224, using the AdamW optimizer with a learning rate of 1e$-$6. 
We apply 8-bit QAT exclusively to the ViT-based backbone, while keeping the rest of the detection architecture in full precision, as only the backbone is implemented in the optical domain where bit-precision constraints apply. The remaining components operate in the electronic domain, which does not impose such limitations. Notably, we improved the performance of the MGNet by increasing the embedding dimension from 192 to 384 and doubling the number of attention heads from 3 to 6. This enhancement enables the network to better capture fine-grained features, particularly benefiting the detection of more complex and small-sized objects in the COCO dataset. As shown in Table \ref{table:coco}, our Opto-ViT achieves comparable or even improved performance over its full-precision counterpart. In object detection, Opto-ViT achieves an AP of 30.53\%, outperforming the baseline ViTDet (30.35\%). Similarly, in instance segmentation, Opto-ViT reaches an AP of 26.95\%, closely matching the baseline's 27.12\%. Notably, the masked version of Opto-ViT maintains performance while slightly improving AP for large objects, by skiping 66\% pixels. 

\noindent\textit{3) Video Object Detection:}
We evaluate video object detection on the ImageNet-VID \cite{russakovsky2015imagenet} validation set, which comprises 639 video sequences covering 30 object categories, with up to 2,895 frames per sequence. 
The ViTDet base model is fine‑tuned using 8‑bit QAT for 50 epochs with a learning rate of 1e${-}$5.
Table \ref{table:vid} reports mean average precision (mAP) along with mAP@0.50 (mAP‑50) and mAP@0.75 (mAP‑75). The full‑precision ViTDet baseline achieves 54.90\% mAP, 80.51\% mAP‑50, and 62.52\% mAP‑75. After QAT, Opto‑ViT attains 53.39\% mAP (1.51\% drop), 79.62\% mAP‑50 (0.89\% drop), and 60.99\% mAP‑75 (1.53\% drop). Adding mask‑guided patch pruning (Opto‑ViT Mask) with 68\% pixel-skip ratio yields a slight further reduction to 53.01\% mAP, 79.12\% mAP‑50, and 59.81\% mAP‑75.
Our Opto‑ViT retain over 96\% of the original detection accuracy, incurring less than a 1.6\% drop in mAP, effectively preserves video detection performance while enabling silicon‑photonic deployment.


\begin{table}[t]
\caption{Top‑1 classification accuracy (\%) of baseline ViT and 8‑bit quantized Opto‑ViT models on CIFAR‑10 and Tiny‑ImageNet.}
\vspace{-.5em}
\centering
\setlength{\tabcolsep}{2.3pt} 
\begin{tabular}{llccc}
\toprule
\textbf{Dataset} & \textbf{Model} & \textbf{Resolution} & \textbf{skip\%} & \textbf{Acc.(ViT / Opto-ViT)} \\
\midrule
\multirow{4}{*}{CIFAR-10} 
    & Tiny     & \multirow{4}{*}{\(96\times96\)} &-  & 97.10\% / 96.56\% \\
    & Small    &                                 &-  & 97.86\% / 97.67\% \\
    & Base     &                                 &-  & 98.56\% / 98.16\% \\
    & Large    &                                 &-  & 98.89\% / 97.87\% \\

\midrule
\multirow{3}{*}{Tiny-ImageNet}             
                      & Base       & \(96\times96\)      &-  & 82.13\% / 80.56\%     \\
                      & Base       & \(224\times224\)    &-  & 85.51\% / 84.64\%    \\
                      & Base Mask  & \(224\times224\)    &0.67  &  -  / 80.12\%    \\
\bottomrule
\end{tabular}
\vspace{-1em}
\label{table:classification}
\end{table}

\begin{table}[t]
\caption{Average precision (AP) metrics for object detection and instance segmentation on the COCO validation set using Mask R-CNN with different ViT-based backbones.}
\vspace{-.5em}
\centering
\setlength{\tabcolsep}{4pt} 
\begin{tabular}{lccccccc}
\toprule
\multirow{2}[7]{*}{\textbf{backbone}} & \multirow{3}{*}{\textbf{\shortstack{skip\%\\(pixel)}}} & \multicolumn{6}{c}{\textbf{Mask R-CNN}}   \\
\cmidrule(l){3-8}
& & \(\text{AP}\) & \(\text{AP}^{50}\) & \(\text{AP}^{75}\) & \(\text{AP}^s\) & \(\text{AP}^m\) & \(\text{AP}^l\) \\
\midrule
\multicolumn{7}{l}{\textit{object detection:}} \\
    ViTDet     &- & 30.35   & 46.98    & 32.24   & 6.89 & 32.48   &55.48 \\
    \textbf{Opto-ViT}   &- & 30.53   & 46.76    & 32.32   & 6.79 & 31.66   &56.58 \\
    \textbf{Opto-ViT Mask}  &0.66 & 30.44   & 46.59    & 32.32   & 6.64 & 31.54   &56.67 \\
\midrule
\multicolumn{7}{l}{\textit{instance segmentation:}} \\
    ViTDet     &- & 27.12 &44.17 &28.26 &3.79 &26.29 &54.60  \\
    \textbf{Opto-ViT}  &-  & 26.95 &43.60 &27.97 &4.09 &25.48 &54.99 \\
    \textbf{Opto-ViT Mask}  &0.66 & 26.85   & 43.37   & 27.93   & 3.90& 25.29   &54.91  \\
\bottomrule
\end{tabular}
\vspace{-1.5em}
\label{table:coco}
\end{table}

\begin{table}[t]
\caption{Mean average precision (mAP) at IoU thresholds for video object detection on the ImageNet-VID validation set using Opto-ViT variants.}
\vspace{-.5em}
\centering
\begin{tabular}{lcccc}
\toprule
  & \textbf{skip\% (pixel)} & mAP & mAP-50 & mAP-75 \\
\midrule
    ViTDet                    &- & 0.549         &  0.8051    &  0.6252    \\
    \textbf{Opto-ViT}       &- & 0.5339     & 0.7962    & 0.6099     \\
    \textbf{Opto-ViT Mask}  &0.68 & 0.5301     & 0.7912    & 0.5981  \\
\bottomrule
\end{tabular}
\vspace{-2em}
\label{table:vid}
\end{table}

\noindent\textbf{Performance Comparison Vs. SiPh Accelerators.} Table \ref{cmp} compares the efficiency of various MR-based optical accelerators, including LightBulb \cite{zokaee2020lightbulb}, HolyLight \cite{liu2019holylight}, HQNNA \cite{sunny2022silicon}, Robin \cite{sunny2021robin}, CrossLight \cite{sunny2021crosslight}, Lightator \cite{morsali2024lightator}, and Opto-ViT. While other designs do not support ViT, we reconstructed each design to closely match the original, leveraging our evaluation framework and proprietary simulator, and ensured a consistent area constraint across all accelerators (approximately 20–60$\,\text{mm}^2$). We observe that Opto-ViT significantly outperforms comparable designs with 100.4 KFPS/W serving as our reference baseline. Relative to this, LightBulb delivers 73.9\% lower efficiency, HolyLight 2941.2\% lower, HQNNA 190.2\% lower, Robin 115.9\% lower, and CrossLight 90.9\% lower at its best. Only Lightator at its best exceeds our work. 

\begin{table}[h]
\centering
\vspace{-1em}
\caption{Comparison with SOTA SiPh accelerators.} \vspace{-1em}
\label{cmp}
\footnotesize
\resizebox{\columnwidth}{!}{%
\begin{tabular}{lccccccc}
\hline
\textbf{Designs} & \textbf{\begin{tabular}[c]{@{}c@{}}LightBulb\\\cite{zokaee2020lightbulb}\end{tabular}} & \textbf{\begin{tabular}[c]{@{}c@{}}HolyLight\\\cite{liu2019holylight}\end{tabular}} & \textbf{\begin{tabular}[c]{@{}c@{}}HQNNA\\\cite{sunny2022silicon}\end{tabular}} & \textbf{\begin{tabular}[c]{@{}c@{}}Robin\\\cite{sunny2021robin}\end{tabular}} & \textbf{\begin{tabular}[c]{@{}c@{}}CrossLight\\\cite{sunny2021crosslight}\end{tabular}} & \textbf{\begin{tabular}[c]{@{}c@{}}Lightator\\\cite{morsali2024lightator}\end{tabular}} & \textbf{Opto‑ViT} \\ \hline
Node (nm)        & 32                                                                                     & 32                                                                                    & 45                                                                                & 45                                                                              & *                                                                                         & 45                                                                                        & 45           \\
KFPS/W           & 57.75                                                                                  & 3.3                                                                                   & 34.6                                                                              & 46.5                                                                            & 10.78–52.59                                                                               & 61.61–188.24                                                                              & 100.4         \\ \hline
Improv.          & 73.9\% (\(\uparrow\))                                                                 & 2941.2\% (\(\uparrow\))                                                               & 190.2\% (\(\uparrow\))                                                            & 115.9\% (\(\uparrow\))                                                          & 90.9\% (\(\uparrow\))                                                                     & –46.7\% (\(\downarrow\))                                                                  & ref           \\ \hline
\end{tabular}
}
\\
\footnotesize{$^{*}$Data not reported/not achievable~\cite{sunny2021crosslight}.}
\end{table}

\noindent\textbf{Performance Comparison Vs. Common Computing Platforms.}
We evaluate the energy efficiency of Opto-ViT against leading-edge inference platforms, including the Xilinx VCK190 FPGA and NVIDIA A100 GPU with TensorRT optimization, following the configurations outlined in \cite{dong2024eq}.
For a fair comparison, all platforms execute the same ViT model using INT8 precision, which is commonly supported in hardware accelerators and inference runtimes. 
The findings underscore the outstanding energy efficiency of Opto-ViT, which achieves two to three orders of magnitude greater efficiency. Specifically, Opto-ViT attains a peak performance of 100.4 KFPS/W, while Xilinx VCK190 and NVIDIA A100 deliver 1.42 and 0.86 KFPS/W, respectively.

\section{Conclusions \& Discussions}
We present Opto-ViT, a novel near-sensor hybrid photonic-electronic accelerator designed to efficiently run Vision Transformer (ViT) models under stringent energy and bandwidth constraints. By combining a silicon-photonic matrix engine—featuring VCSEL-driven optical inputs and microring-resonator-based processing—with lightweight region-of-interest (RoI) pruning, Opto-ViT enables patch-level sparsity and energy-aware inference at the sensor edge. Our accelerator delivers 100.4 KFPS/W and achieves up to 84\% energy savings, while maintaining less than 1.6\% accuracy loss across image, video, and object detection tasks.

This work highlights the viability of photonic computation for deploying complex transformer architectures in always-on, edge-AI environments. Looking forward, our framework opens new research avenues in adaptive photonic reconfiguration, multi-modal sensor fusion, and programmable optics to further optimize ViT-like models for diverse edge applications. Moreover, co-designing future models with hardware constraints in mind—particularly around bandwidth and tuning latency—can unlock even broader applicability for hybrid photonic systems.

\section*{Acknowledgments}
\small{This work is supported in part by Semiconductor Research Corporation (SRC) and the National Science Foundation (NSF) under grant no. 2216772, 2228028, 2046226, and 2006788.}

\small\bibliographystyle{IEEEtran}
\bibliography{IEEEabrv,./Ref}\vspace{-2em}

\begin{thebibliography}{10}
\providecommand{\url}[1]{#1}
\csname url@samestyle\endcsname
\providecommand{\newblock}{\relax}
\providecommand{\bibinfo}[2]{#2}
\providecommand{\BIBentrySTDinterwordspacing}{\spaceskip=0pt\relax}
\providecommand{\BIBentryALTinterwordstretchfactor}{4}
\providecommand{\BIBentryALTinterwordspacing}{\spaceskip=\fontdimen2\font plus
\BIBentryALTinterwordstretchfactor\fontdimen3\font minus \fontdimen4\font\relax}
\providecommand{\BIBforeignlanguage}[2]{{%
\expandafter\ifx\csname l@#1\endcsname\relax
\typeout{** WARNING: IEEEtran.bst: No hyphenation pattern has been}%
\typeout{** loaded for the language `#1'. Using the pattern for}%
\typeout{** the default language instead.}%
\else
\language=\csname l@#1\endcsname
\fi
#2}}
\providecommand{\BIBdecl}{\relax}
\BIBdecl

\bibitem{carey2013100}
S.~J. Carey, A.~Lopich, D.~R. Barr, B.~Wang, and P.~Dudek, ``A 100,000 fps vision sensor with embedded 535gops/w 256$\times$ 256 simd processor array,'' in \emph{Symposium on VLSI}.\hskip 1em plus 0.5em minus 0.4em\relax IEEE, 2013.

\bibitem{hsu20200}
T.-H. Hsu, Y.-R. Chen, R.-S. Liu, C.-C. Lo, K.-T. Tang, M.-F. Chang, and C.-C. Hsieh, ``A 0.5-v real-time computational cmos image sensor with programmable kernel for feature extraction,'' \emph{IEEE JSSC}, vol.~56, pp. 1588--1596, 2020.

\bibitem{yamazaki20174}
T.~Yamazaki \emph{et~al.}, ``4.9 a 1ms high-speed vision chip with 3d-stacked 140gops column-parallel pes for spatio-temporal image processing,'' in \emph{2017 IEEE International Solid-State Circuits Conference (ISSCC)}.\hskip 1em plus 0.5em minus 0.4em\relax IEEE, 2017, pp. 82--83.

\bibitem{angizi2018cmp}
S.~Angizi, Z.~He, A.~S. Rakin, and D.~Fan, ``Cmp-pim: an energy-efficient comparator-based processing-in-memory neural network accelerator,'' in \emph{Proceedings of the 55th Annual Design Automation Conference}, 2018, pp. 1--6.

\bibitem{morsali2023deep}
M.~Morsali, S.~Tabrizchi, M.~Liehr, N.~Cady, M.~Imani, A.~Roohi, and S.~Angizi, ``Deep mapper: A multi-channel single-cycle near-sensor dnn accelerator,'' in \emph{2023 IEEE International Conference on Rebooting Computing (ICRC)}.\hskip 1em plus 0.5em minus 0.4em\relax IEEE, 2023, pp. 1--5.

\bibitem{xu2020macsen}
H.~Xu, Z.~Li, N.~Lin, Q.~Wei, F.~Qiao, X.~Yin, and H.~Yang, ``Macsen: A processing-in-sensor architecture integrating mac operations into image sensor for ultra-low-power bnn-based intelligent visual perception,'' \emph{IEEE TCAS II}, vol.~68, pp. 627--631, 2020.

\bibitem{xu2021senputing}
H.~Xu \emph{et~al.}, ``Senputing: An ultra-low-power always-on vision perception chip featuring the deep fusion of sensing and computing,'' \emph{IEEE TCASI}, 2021.

\bibitem{tabrizchi2023appcip}
S.~Tabrizchi \emph{et~al.}, ``Appcip: Energy-efficient approximate convolution-in-pixel scheme for neural network acceleration,'' \emph{IEEE JETCAS}, pp. 225--236, 2023.

\bibitem{angizi2023pisa}
S.~Angizi \emph{et~al.}, ``Pisa: A non-volatile processing-in-sensor accelerator for imaging systems,'' \emph{IEEE TETC}, 2023.

\bibitem{abedin2022mr}
M.~Abedin \emph{et~al.}, ``Mr-pipa: An integrated multi-level rram (hfo x) based processing-in-pixel accelerator,'' \emph{IEEE JXCDC}, 2022.

\bibitem{tabrizchi2024pinsim}
S.~Tabrizchi, M.~Morsali, D.~Pan, S.~Angizi, and A.~Roohi, ``Pinsim: A processing in-and near-sensor simulator to model intelligent vision sensors,'' \emph{IEEE Computer Architecture Letters}, 2024.

\bibitem{tabrizchi2024apris}
S.~Tabrizchi, R.~Gaire, M.~Morsali, M.~Liehr, N.~Cady, S.~Angizi, and A.~Roohi, ``Apris: Approximate processing reram in-sensor architecture enabling artificial-intelligence-powered edge,'' \emph{IEEE Transactions on Emerging Topics in Computing}, 2024.

\bibitem{najafi2024enabling}
D.~Najafi, M.~Morsali, R.~Zhou, A.~Roohi, A.~Marshall, D.~Misra, and S.~Angizi, ``Enabling normally-off in situ computing with a magneto-electric fet-based sram design,'' \emph{IEEE Transactions on Electron Devices}, vol.~71, no.~4, pp. 2742--2748, 2024.

\bibitem{el1999pixel}
A.~E. Gamal, D.~X.~D. Yang, and B.~A. Fowler, ``Pixel-level processing: why, what, and how?'' in \emph{Sensors, Cameras, and Applications for Digital Photography}, vol. 3650.\hskip 1em plus 0.5em minus 0.4em\relax SPIE, 1999, pp. 2--13.

\bibitem{song2022reconfigurable}
R.~Song, K.~Huang, Z.~Wang, and H.~Shen, ``A reconfigurable convolution-in-pixel cmos image sensor architecture,'' \emph{IEEE TCSVT}, 2022.

\bibitem{roohi2023pipsim}
A.~Roohi, S.~Tabrizchi, M.~Morsali, D.~Z. Pan, and S.~Angizi, ``Pipsim: A behavior-level modeling tool for cnn processing-in-pixel accelerators,'' \emph{IEEE Transactions on Computer-Aided Design of Integrated Circuits and Systems}, vol.~43, no.~1, pp. 141--150, 2023.

\bibitem{attention}
A.~Vaswani \emph{et~al.}, ``Attention is all you need,'' in \emph{Advances in Neural Information Processing Systems}, I.~Guyon \emph{et~al.}, Eds., vol.~30.\hskip 1em plus 0.5em minus 0.4em\relax Curran Associates, Inc., 2017.

\bibitem{Caron_2021_ICCV}
M.~Caron, H.~Touvron, I.~Misra, H.~J\'egou, J.~Mairal, P.~Bojanowski, and A.~Joulin, ``Emerging properties in self-supervised vision transformers,'' in \emph{Proceedings of the IEEE/CVF International Conference on Computer Vision (ICCV)}, October 2021, pp. 9650--9660.

\bibitem{ahmed2025deepcompress}
S.~Ahmed, A.~Al~Arafat, D.~Najafi, A.~Mahmood, M.~N. Rizve, M.~Al~Nahian, R.~Zhou, S.~Angizi, and A.~S. Rakin, ``Deepcompress-vit: Rethinking model compression to enhance efficiency of vision transformers at the edge,'' in \emph{Proceedings of the Computer Vision and Pattern Recognition Conference}, 2025, pp. 30\,147--30\,156.

\bibitem{vita2023}
S.~Nag, G.~Datta, S.~Kundu, N.~Chandrachoodan, and P.~A. Beerel, ``Vita: A vision transformer inference accelerator for edge applications,'' in \emph{2023 IEEE International Symposium on Circuits and Systems (ISCAS)}.\hskip 1em plus 0.5em minus 0.4em\relax IEEE, 2023, pp. 1--5.

\bibitem{retransformer}
X.~Yang, B.~Yan, H.~Li, and Y.~Chen, ``Retransformer: Reram-based processing-in-memory architecture for transformer acceleration,'' in \emph{Proceedings of the 39th International Conference on Computer-Aided Design}, 2020, pp. 1--9.

\bibitem{tron2023afifi}
S.~Afifi, F.~Sunny, M.~Nikdast, and S.~Pasricha, ``Tron: Transformer neural network acceleration with non-coherent silicon photonics,'' in \emph{Proceedings of the Great Lakes Symposium on VLSI 2023}, 2023, pp. 15--21.

\bibitem{choi2015energy}
J.~Choi, S.~Park, J.~Cho, and E.~Yoon, ``An energy/illumination-adaptive cmos image sensor with reconfigurable modes of operations,'' \emph{IEEE Journal of Solid-State Circuits}, vol.~50, no.~6, pp. 1438--1450, 2015.

\bibitem{hsu2019ai}
T.-H. Hsu \emph{et~al.}, ``{AI} edge devices using computing-in-memory and processing-in-sensor: from system to device,'' in \emph{IEDM}, 2019.

\bibitem{angizi2023near}
S.~Angizi, M.~Morsali, S.~Tabrizchi, and A.~Roohi, ``A near-sensor processing accelerator for approximate local binary pattern networks,'' \emph{IEEE Transactions on Emerging Topics in Computing}, vol.~12, no.~1, pp. 73--83, 2023.

\bibitem{sunny2021robin}
F.~P. Sunny, A.~Mirza, M.~Nikdast, and S.~Pasricha, ``Robin: A robust optical binary neural network accelerator,'' \emph{ACM TECS}, no.~5s, pp. 1--24, 2021.

\bibitem{najafi2025neuro}
D.~Najafi, H.~E. Barkam, M.~Morsali, S.~Jeong, T.~Das, A.~Roohi, M.~Nikdast, M.~Imani, and S.~Angizi, ``Neuro-photonix: Enabling near-sensor neuro-symbolic ai computing on silicon photonics substrate,'' \emph{IEEE Transactions on Circuits and Systems for Artificial Intelligence}, 2025.

\bibitem{sunny2021crosslight}
F.~Sunny, A.~Mirza, M.~Nikdast, and S.~Pasricha, ``Crosslight: A cross-layer optimized silicon photonic neural network accelerator,'' in \emph{DAC}.\hskip 1em plus 0.5em minus 0.4em\relax IEEE, 2021, pp. 1069--1074.

\bibitem{cheng2020silicon}
Q.~Cheng \emph{et~al.}, ``Silicon photonics codesign for deep learning,'' \emph{Proceedings of the IEEE}, vol. 108, pp. 1261--1282, 2020.

\bibitem{morsali2024oisa}
M.~Morsali, S.~Tabrizchi, D.~Najafi, M.~Imani, M.~Nikdast, A.~Roohi, and S.~Angizi, ``Oisa: Architecting an optical in-sensor accelerator for efficient visual computing,'' in \emph{2024 Design, Automation \& Test in Europe Conference \& Exhibition (DATE)}.\hskip 1em plus 0.5em minus 0.4em\relax IEEE, 2024, pp. 1--6.

\bibitem{dosovitskiy2021an}
\BIBentryALTinterwordspacing
A.~Dosovitskiy \emph{et~al.}, ``An image is worth 16x16 words: Transformers for image recognition at scale,'' in \emph{International Conference on Learning Representations}, 2021. [Online]. Available: \url{https://openreview.net/forum?id=YicbFdNTTy}
\BIBentrySTDinterwordspacing

\bibitem{devlin2019bertpretrainingdeepbidirectional}
J.~Devlin, M.-W. Chang, K.~Lee, and K.~Toutanova, ``Bert: Pre-training of deep bidirectional transformers for language understanding,'' \emph{arXiv preprint arXiv:1810.04805}, 2019.

\bibitem{liu2019holylight}
W.~Liu, W.~Liu, Y.~Ye, Q.~Lou, Y.~Xie, and L.~Jiang, ``Holylight: A nanophotonic accelerator for deep learning in data centers,'' in \emph{DATE}.\hskip 1em plus 0.5em minus 0.4em\relax IEEE, 2019, pp. 1483--1488.

\bibitem{zokaee2020lightbulb}
F.~Zokaee, Q.~Lou, N.~Youngblood, W.~Liu, Y.~Xie, and L.~Jiang, ``Lightbulb: A photonic-nonvolatile-memory-based accelerator for binarized convolutional neural networks,'' in \emph{DATE}.\hskip 1em plus 0.5em minus 0.4em\relax IEEE, 2020, pp. 1438--1443.

\bibitem{zhao2019hardware}
Z.~Zhao, D.~Liu, M.~Li, Z.~Ying, L.~Zhang, B.~Xu, B.~Yu, R.~T. Chen, and D.~Z. Pan, ``Hardware-software co-design of slimmed optical neural networks,'' in \emph{ASP-DAC}.\hskip 1em plus 0.5em minus 0.4em\relax IEEE, 2019, pp. 705--710.

\bibitem{morsali2024lightator}
M.~Morsali, B.~Reidy, D.~Najafi, S.~Tabrizchi, M.~Imani, M.~Nikdast, A.~Roohi, R.~Zand, and S.~Angizi, ``Lightator: An optical near-sensor accelerator with compressive acquisition enabling versatile image processing,'' \emph{arXiv preprint arXiv:2403.05037}, 2024.

\bibitem{bogaerts2012silicon}
W.~Bogaerts, P.~De~Heyn, T.~Van~Vaerenbergh, K.~De~Vos, S.~Kumar~Selvaraja, T.~Claes, P.~Dumon, P.~Bienstman, D.~Van~Thourhout, and R.~Baets, ``Silicon microring resonators,'' \emph{Laser \& Photonics Reviews}, pp. 47--73, 2012.

\bibitem{softmaxgelu}
C.~Peltekis, K.~Alexandridis, and G.~Dimitrakopoulos, ``Reusing softmax hardware unit for gelu computation in transformers,'' in \emph{2024 IEEE 6th International Conference on AI Circuits and Systems (AICAS)}, 2024, pp. 159--163.

\bibitem{NCSU_PDK}
\BIBentryALTinterwordspacing
(2011) Ncsu eda freepdk45. [Online]. Available: \url{http://www.eda.ncsu.edu/wiki/FreePDK45}
\BIBentrySTDinterwordspacing

\bibitem{DC}
{Synopsys, Inc.}, ``Synopsys design compiler, product version 14.9.2014,'' 2014.

\bibitem{duong2014case}
L.~H.~K. Duong, M.~Nikdast, S.~Le~Beux, J.~Xu, X.~Wu, Z.~Wang, and P.~Yang, ``A case study of signal-to-noise ratio in ring-based optical networks-on-chip,'' \emph{IEEE Design \& Test}, vol.~31, 2014.

\bibitem{kaiser2024energy}
M.~A.-A. Kaiser, S.~Sarkar, P.~A. Beerel, A.~R. Jaiswal, and G.~Datta, ``Energy-efficient \& real-time computer vision with intelligent skipping via reconfigurable {CMOS} image sensors,'' \emph{arXiv preprint arXiv:2409.17341}, 2024.

\bibitem{jacob2018quantization}
B.~Jacob, S.~Kligys, B.~Chen, M.~Zhu, M.~Tang, A.~Howard, H.~Adam, and D.~Kalenichenko, ``Quantization and training of neural networks for efficient integer-arithmetic-only inference,'' in \emph{Proceedings of the IEEE conference on computer vision and pattern recognition}, 2018, pp. 2704--2713.

\bibitem{bengio2013estimating}
Y.~Bengio, N.~L{\'e}onard, and A.~Courville, ``Estimating or propagating gradients through stochastic neurons for conditional computation,'' \emph{arXiv preprint arXiv:1308.3432}, 2013.

\bibitem{li2023vit}
Z.~Li and Q.~Gu, ``I-vit: Integer-only quantization for efficient vision transformer inference,'' in \emph{Proceedings of the IEEE/CVF International Conference on Computer Vision}, 2023, pp. 17\,065--17\,075.

\bibitem{ridnik2021imagenet}
T.~Ridnik, E.~Ben-Baruch, A.~Noy, and L.~Zelnik-Manor, ``Imagenet-21k pretraining for the masses,'' \emph{arXiv preprint arXiv:2104.10972}, 2021.

\bibitem{CIFAR10}
A.~Krizhevsky and G.~Hinton, ``Convolutional deep belief networks on cifar-10,'' \emph{Unpublished manuscript}, vol.~40, no.~7, pp. 1--9, 2010.

\bibitem{deng2009imagenet}
J.~Deng, W.~Dong, R.~Socher, L.~Li, K.~Li, and L.~Fei-Fei, ``Imagenet: A large-scale hierarchical image database,'' in \emph{Proc. Conference on Computer Vision and Pattern Recognition}, 2009, pp. 248--255.

\bibitem{lin2014microsoft}
T.-Y. Lin, M.~Maire, S.~Belongie, J.~Hays, P.~Perona, D.~Ramanan, P.~Doll{\'a}r, and C.~L. Zitnick, ``Microsoft coco: Common objects in context,'' in \emph{Computer vision--ECCV 2014: 13th European conference, zurich, Switzerland, September 6-12, 2014, proceedings, part v 13}.\hskip 1em plus 0.5em minus 0.4em\relax Springer, 2014, pp. 740--755.

\bibitem{li2022exploring}
Y.~Li, H.~Mao, R.~Girshick, and K.~He, ``Exploring plain vision transformer backbones for object detection,'' in \emph{European conference on computer vision}.\hskip 1em plus 0.5em minus 0.4em\relax Springer, 2022, pp. 280--296.

\bibitem{he2017mask}
K.~He, G.~Gkioxari, P.~Doll{\'a}r, and R.~Girshick, ``Mask r-cnn,'' in \emph{Proceedings of the IEEE international conference on computer vision}, 2017, pp. 2961--2969.

\bibitem{russakovsky2015imagenet}
O.~Russakovsky, J.~Deng, H.~Su, J.~Krause, S.~Satheesh, S.~Ma, Z.~Huang, A.~Karpathy, A.~Khosla, M.~Bernstein \emph{et~al.}, ``Imagenet large scale visual recognition challenge,'' \emph{International journal of computer vision}, vol. 115, pp. 211--252, 2015.

\bibitem{sunny2022silicon}
F.~Sunny, M.~Nikdast, and S.~Pasricha, ``A silicon photonic accelerator for convolutional neural networks with heterogeneous quantization,'' in \emph{GLSVLSI}, 2022, pp. 367--371.

\bibitem{dong2024eq}
P.~Dong \emph{et~al.}, ``Eq-vit: Algorithm-hardware co-design for end-to-end acceleration of real-time vision transformer inference on versal acap architecture,'' \emph{IEEE Transactions on Computer-Aided Design of Integrated Circuits and Systems}, vol.~43, no.~11, pp. 3949--3960, 2024.

\end{thebibliography}

\end{document}